\documentclass[12pt]{article}
\usepackage[dvips]{epsfig}
\usepackage{graphicx}
\usepackage{amssymb,amsmath,bm}
\usepackage{setspace}
\usepackage{enumitem}
\usepackage{mdframed}
\usepackage[flushleft]{threeparttable}
\usepackage{algorithm}
\usepackage{algpseudocode}

\usepackage{hyperref}

\usepackage[T1]{fontenc}
\usepackage{tabularx,ragged2e,booktabs,caption}
\usepackage{array}
\newcolumntype{L}[1]{>{\raggedright\let\newline\\\arraybackslash\hspace{0pt}}m{#1}}
\newcolumntype{C}[1]{>{\centering\let\newline\\\arraybackslash\hspace{0pt}}m{#1}}
\newcolumntype{R}[1]{>{\raggedleft\let\newline\\\arraybackslash\hspace{0pt}}m{#1}}

\usepackage[english]{babel}
\usepackage[round,sort&compress]{natbib}
\usepackage{multirow}
\usepackage{theorem}
\usepackage{float}

\def\00{\mathrm{0}}

\begin{document}
	
	\thispagestyle{empty} \baselineskip=28pt \vskip 5mm
	\begin{center} {\LARGE{\bf Functional Outlier Detection and Taxonomy by Sequential Transformations}}
	\end{center}
	
	\baselineskip=12pt \vskip 10mm

	\begin{center}\large
		Wenlin Dai\footnote[1]{\baselineskip=10pt Institute of Statistics and Big Data, Renmin University of China, Beijing 100872, China. E--mail: wenlin.dai@ruc.edu.cn}, 
		Tom\'a\v s Mrkvi\v cka\footnote[2]{Department of Applied Mathematics and Informatics, Faculty of Economics, University of South Bohemia, Studentsk{\'a} 13, 37005 \v{C}esk\'e Bud\v{e}jovice, Czech Republic. E--mail: mrkvicka.toma@gmail.com}, 
		Ying Sun\footnote[3]{\baselineskip=10pt Statistics Program,
			King Abdullah University of Science and Technology,
			Thuwal 23955-6900, Saudi Arabia. E--mail: ying.sun@kaust.edu.sa, marc.genton@kaust.edu.sa \\This research was supported by the King Abdullah University of Science and Technology (KAUST).}, and Marc G. Genton$^3$
\end{center}

\baselineskip=17pt \vskip 10mm \centerline{\today} \vskip 15mm

%%%%%%%%%%%%%%%%%%%%%%%%%%%%%%%%%%%%%%%%%%%%%%%%%%%%%%%%%%%%%%%%%%%%%%%%
\begin{center}
	{\large{\bf Abstract}}
\end{center}
Functional data analysis can be seriously impaired by abnormal observations, which can be classified as either magnitude or shape outliers based on their way of deviating from the bulk of data. 
Identifying magnitude outliers is relatively easy, while detecting shape outliers is much more challenging. 
We propose turning the shape outliers into magnitude outliers through data transformation and detecting them using the functional boxplot. 
Besides easing the detection procedure, applying several transformations sequentially provides a reasonable taxonomy for the flagged outliers. 
A joint functional ranking, which consists of several transformations, is also defined here.
Simulation studies are carried out to evaluate the performance of the proposed method using different functional depth notions. 
Interesting results are obtained in several practical applications.

\baselineskip=14pt

\par\vfill\noindent
\par\medskip\noindent
{\bf Short title:} Functional Outlier Detection and Taxonomy

\clearpage\pagebreak\newpage \pagenumbering{arabic}
\baselineskip=26pt

\vskip 12pt
\section{Introduction}
Functional data analysis is attracting growing attention as data are increasingly recorded as curves, images, or tensors. Ever since the founding work of \citet{ramsayfunctional}, a large body of literature has emerged on different perspectives of functional data analysis, e.g., nonparametric methods \citep{ferraty2006nonparametric}, statistical inference \citep{horvath2012inference}, and regression models \citep{yao2005functional}. We refer the readers to \citet{wang2016functional} for a comprehensive review.

Functional data analysis can be severely biased if the data are contaminated by abnormal observations.
Therefore, it is necessary to reduce the influence of contamination and to analyze the data robustly. 
Data ranking is popularly implemented to provide robust descriptions of point-type data. 
Univariate data are naturally sorted monotonically; multivariate data, lacking a natural order, are commonly sorted from the center outward using a measure of statistical depth or outlyingness. 
During the past decade, statistical depth was generalized to the functional domain as a tool to measure the centrality of functional data. 
Various functional depth notions have been investigated in the literature; see Section~3.1 for more detail.
These notions can be divided into two subclasses, integrated and non-integrated, based on their definitions; regarding the type of utilized information, they can be classified as rank-based or distance-based.

\begin{figure}[!b]
	\centering
	\includegraphics[width=\textwidth]{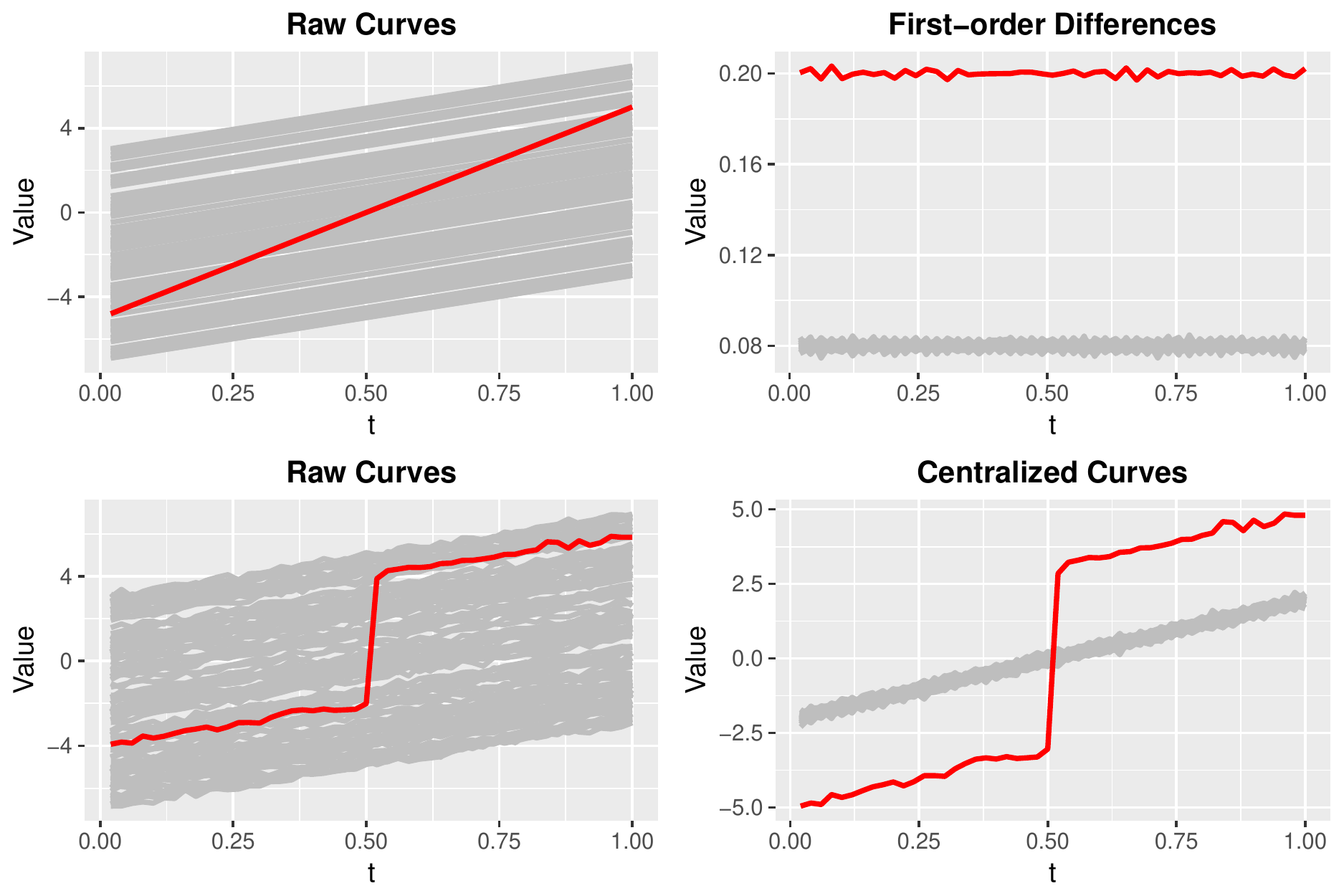}\\
	\caption{Shape outliers can be changed into magnitude outliers through some type of transformation. Top panels: a slope shape outlier (the red line) is changed into a magnitude outlier by taking the first-order differences of the raw curves; bottom panels: a jump shape outlier (the red curve) is changed into a magnitude outlier by shifting each raw curve so its mean value becomes zero.}
	\label{motivation}
\end{figure}

Abnormal observations, called functional outliers, commonly fall into two categories: magnitude outliers and shape outliers. 
An observation is regarded as a {\em magnitude outlier} if it is outlying in some part or across the whole design domain. It is viewed as a shape outlier if it has a different shape from the bulk of data, even though it may not be outlying throughout the entire interval.
Magnitude outliers can be detected and visualized well by existing exploratory methods, e.g., the functional bagplot \citep{hyndman2010rainbow}, the functional boxplot \citep{sun2011functional,sun2012adjusted}, and the global envelope \citep{myllymaki2017}. 
Shape outliers, on the other hand, are much more challenging to handle  \citep{hubert2015multivariate,dai2019directional,nagy2017depth}. 
To tackle the shape outliers, some researchers proposed decomposing the overall functional depth (or outlyingness) into just magnitude and shape depth (or outlyingness) in order to capture the shape outliers more accurately. 
Examples include the outliergram \citep{arribas2014shape}, the functional outlier map \citep{rousseeuw2016measure}, the total variation depth \citep{huang2019decomposition}, and the magnitude-shape plot \citep{dai2018msplot}. 
Researchers also defined depth notions that utilize the local geometric features of the curves \citep{kuhnt2016angle,nagy2017depth}. The distribution of the resulting functional depth values is usually unknown; as a result, it is difficult to choose an accurate cutoff value for detecting outliers.

In this paper, we show that most commonly encountered functional shape outliers can be transformed into magnitude outliers, which are easier to recognize with some popular diagnostics tools.
For instance, by taking the first-order differences or derivatives of the raw curve, we may change a shape outlier with an anomalous slope into a magnitude outlier (see the top panels of Figure~\ref{motivation}). Curve transformation turns out to be an effective way to improve the performance of outlier detection methods in recognizing shape outliers. 
Another advantage is that the plot of the transformed curves together with the type of transformation provides an intuitive graphical interpretation to the specific mechanism of outlyingness. 
We also show that the central region not only accurately describes the pattern of data but also visualizes the location of the detected anomaly, when it is constructed with some specific depth, e.g., $L^{\infty}$ depth \citep{long2015study} and extreme rank length depth \citep{myllymaki2017}. 
Moreover, various transformations focus on different perspectives of curves so we may apply a sequence of transformations in a row and classify the outliers with respect to the transformations.

To detect magnitude outliers, we choose the functional boxplot \citep{sun2011functional}, which is a graphical tool popularly utilized for functional exploratory data analysis, mimicking Tukey's boxplot for scalars. The functional boxplot is constructed by ordering a group of univariate curves from the center outward according to the modified band depth \citep{lopez2009concept}, or any other user-specified depth notions.
The envelope of the $50\%$ deepest curves forms the $50\%$ central region; by inflating this region by $F^*$ times its vertical range, two fences are obtained for the detection of outliers. The default value is set to $F^*=1.5$, which can be also adaptively determined by considering the underlying correlation structure \citep{sun2012adjusted}.
Eventually, the envelope of the central $50\%$ region, the median curve, and the maximum non-outlying envelope are used as descriptive statistics; functional outliers, if detected, are also visualized. 

Besides outlier detection, another problem that could benefit from data transformation is functional testing, e.g., spatial point process model testing \citep{myllymaki2017} and spatial covariance function properties assessment \citep{huang2018visualization}, where one curve is examined through functional replicates generated from the model under the null hypothesis. The testing curve could be abnormal due to either magnitude or shape, which leads to the same challenge as the outlier detection problem.
\citet{myllymaki2017} proposed a global envelope test, which sorts the tested curve together with the simulated curves according to their depth values. Hence, an accurate ranking is critical for the construction of this tool. 
We illustrate that merging the ranking results from different transformations into a joint functional order leads to a better ranking overall for the testing problem.

The remainder of the paper is organized as follows. In Section~2, we systematically explain the procedure of shape outlier detection based on data transformation and the functional boxplot, and provide some simple and effective transformations according to our numerical studies.
In Sections~3 and 4, we evaluate the effectiveness of the transformations, and search for the proper depth notions to construct the functional boxplot, with two typical outlier detection problems, using Monte Carlo simulations. In Section~5, we apply the proposed method to several datasets and follow with a conclusion in Section~6.

\vskip 12pt
\section{Curve Transformations}
\vskip 5pt
\subsection{Sequential Transformations for Functional Outlier Detection}
We consider a group of functional observations, $X_i\in \mathcal{C}(\mathcal{I})$, $i=1,\dots,n$, where $\mathcal{I}$ is an interval in $\mathbb{R}$, and $\mathcal{C}(\mathcal{I})$ denotes the space of continuous functions defined on $\mathcal{I}$. Assume that $X_i$ follows a distribution defined on $\mathcal{C}(\mathcal{I})$, denoted by $F_X$. For each fixed design point $t_0$ in $\mathcal{I}$, the marginal distribution of $X_i(t_0)$ is denoted by $F_{X(t_0)}$.

As aforementioned, we propose to turn shape outliers into magnitude outliers through some curve-transformation in order to identify the outliers more easily. Denote with $\mathcal{G}$ a transformation defined on $\mathcal{C}(\mathcal{I})$ and with $X_{\rm so}$ a shape outlier with respect to $F_X$. 
For a clean dataset, $\{X_i\}_{i=1}^n$, from the distribution $F_X$, $\{\mathcal{G}(X_i)\}_{i=1}^n$ follows an identical distribution denoted by $F_{\mathcal{G}(X)}$. In the presence of $X_{\rm so}$, $\mathcal{G}(X_{\rm so})$ may be a magnitude outlier rather than a shape outlier with respect to $F_{\mathcal{G}(X)}$. We formalize the whole outlier detection procedure in Algorithm 1: 

\begin{algorithm}
	\begin{algorithmic}
		\State {\bf Algorithm 1: \\Functional Outlier Detection via Sequential Data Transformation}
		\vskip 5pt
		\State 0. Identify outliers from $\{X_i\}_{i=1}^n$ using an effective method, e.g., the functional boxplot constructed with some depth notion $d$; the detected outliers, denoted by $S_0$, are called {\bf $\mathcal{G}_0$--outliers (magnitude outliers)};
		\vskip 5pt
		\State 1. Apply a transformation, $\mathcal{G}_1$, to $\{X_i\}_{i=1}^n$ and get $\{\mathcal{G}_1(X_i)\}_{i=1}^n$;
		\vskip 5pt
		\State 2. Repeat Step 0 on $\{\mathcal{G}_1(X_i)\}_{i=1}^n$; the detected outliers are denoted by $S_1$;  $S_1\setminus S_{0}$ are called {\bf $\mathcal{G}_1$--shape outliers};
		\vskip 5pt 
		\State 3. Apply another transformation, $\mathcal{G}_2$, to $\{\mathcal{G}_1(X_i)\}_{i=1}^n$ when a sequence of transformations is considered and get $\{\mathcal{G}_2\circ\mathcal{G}_1(X_i)\}_{i=1}^n$;
		\vskip 5pt
		\State 4. Repeat Step 0 on $\{\mathcal{G}_2\circ\mathcal{G}_1(X_i)\}_{i=1}^n$; the detected outliers are denoted by $S_2$;  $S_2\setminus (S_{0}\cup S_{1})$ are called {\bf $\mathcal{G}_2\circ\mathcal{G}_1$--shape outliers};
		\vskip 5pt
		\State 5. Repeat Steps 3 and 4 recursively if more types of transformations are considered.
	\end{algorithmic}
\end{algorithm}
The transformation $\mathcal{G}$ can be quite general, involving various types of transforms. For example, it can be transforming a curve into a scalar (functional depth/outlyingness), shifting each curve by its mean value so that the curves achieve the same level, or representing the curves in the frequency domain. Hereafter, we restrict the curve transformation to be similar to the second type, i.e., mappings from $\mathcal{C}(\mathcal{I})$ to $\mathcal{C}(\mathcal{I})$. More examples of transformations are introduced in Section 2.3. As described, if multiple transformations are applied sequentially, we may  simultaneously detect not only the magnitude and shape outliers, and, but also get the taxonomy of the detected outliers. 

\subsection{Sequential Transformations for Functional Testing}
Other than detecting outliers from a given dataset, the data transformations mentioned above are also useful for functional testing problems such as those mentioned in Section 1. For a hypothesis test based on data transformation, we formalize the procedure in Algorithm~2:
\begin{algorithm}
	\begin{algorithmic}
		\State {\bf Algorithm 2: \\Functional Testing via Sequential Data Transformation}
		\vskip 5pt
		\State 0. Generate replicates of the data under the null hypothesis;
		\vskip 5pt
		\State 1. Apply a sequence of transformations, $\{\mathcal{G}_k\}$, $k=0,1,2,\dots$, to the raw data;
		\vskip 5pt
		\State 2. Sort the raw and transformed data respectively, and get the vector of ranks for each observation as $\mathbf{r}_i=(r_{i,0},r_{i,1},r_{i,2},\dots)^{\rm T}$, $i=1,\dots,n$, where $r_{i,k}$ is the rank of $\mathcal{G}_k(X_i)$ among $\{\mathcal{G}_k(X_i)\}_{i=1}^n$, $k=0,1,2,\dots$
		\vskip 5pt
		\State 3. Sort the vectors of rank according to a one-side depth notion, e.g., extreme rank length depth or directional quantile, which will be introduced in Sections 3.15 and 3.1.6;
		\vskip 5pt 
		\State 4. Construct the global envelope test using the ranking results from Step 3.
	\end{algorithmic}
\end{algorithm}
\vskip -10pt
\noindent
Note that Steps 2, 3, and 4 define a joint functional ranking, according to which the global envelope test is performed. 

Transforming the shape outliers into magnitude outliers is the most critical step of the proposed procedures.
It is also important to choose an appropriate depth notion to rank the functional data and, hence, construct the global envelope (or the functional boxplot) that detects the magnitude outliers effectively.

\subsection{Examples of Transformations}
Here, we mention a sequence of simple transformations that are very effective according to our numerical studies. The preliminary step, $\mathcal{T}_0$, is to apply the functional boxplot to the raw curves and define the $\mathcal{T}_0$-outliers as magnitude outliers.
The first transformation, denoted by $\mathcal{T}_1$, shifts each curve $X(t)$ to its center: 
$$\mathcal{T}_1(X)(t)=X(t)-{1\over \lambda(\mathcal{I})}\int_{\mathcal{I}} X(t){\rm d}t,$$
where $\lambda(\mathcal{I})$ is the Lebesgue measure of the interval $\mathcal{I}$. 
$\mathcal{T}_1$ vertically aligns the curves so that their mean values all become zero. After the $\mathcal{T}_1$ transformation, the outliers detected by the functional boxplot usually reveal either local or global abnormal amplitudes. Therefore, $\mathcal{T}_1$-shape outliers can be regarded as amplitude outliers. The second transformation, denoted by $\mathcal{T}_2$, normalizes the centered curves from $\mathcal{T}_1$ with their $L_2$ norms, i.e., 
$$\mathcal{T}_2(X)(t)=\mathcal{T}_1(X)(t)\|\mathcal{T}_1(X)\|_2^{-1},$$
where $\|\mathcal{T}_1(X)\|_2=\left[\int_{\mathcal{I}}\left\{\mathcal{T}_1(X)(t)\right\}^2{\rm d}t\right]^{1/2}$. $\mathcal{T}_2$ filters out the information about both the magnitude and amplitude,  leaving only pure information about the pattern of the raw curves. Thus, $\mathcal{T}_2\circ\mathcal{T}_1$-shape outliers are called pattern outliers.

Another possible sequence of transformations involves taking derivatives or differences of the raw curves. In this sequence, the preliminary step of applying the functional boxplot to the raw curves is the same, denoted here as $\mathcal{D}_0$. The $\mathcal{D}_0$-outliers are also magnitude outliers. As the first step, $\mathcal{D}_1$, we take the first-order derivative or differences of the raw curves and the $\mathcal{D}_1$-shape outliers are called first-order outliers, which indicate abnormal slopes. For the second step, $\mathcal{D}_2$, we take the second-order derivative or differences of the raw curves. These $\mathcal{D}_2\circ\mathcal{D}_1$-shape outliers are called second-order outliers and indicate abnormal curvature. 

Besides the above sequences, we found several other single-step transformations that are useful in some special scenarios. For example, aligning the important features of curves eliminates the phase variation so that the shape outliers can be detected more effectively. This transformation, denoted by $\mathcal{R}$, is expressed as
$$\mathcal{R}(X)(t)=X(r(t)),$$
where $r(t)$ is the warping function on the design interval $\mathcal{I}$.

When the response at each point is multi-dimensional, i.e., multivariate functional data, the abnormal interactions among responses other than the marginal outliers are also interesting to investigate \citep{claeskens2014multivariate}. 
To tackle this challenge, we could calculate the outlyingness of the multivariate functional data at each time point to obtain a univariate curve of outlyingness. This transformation, denoted by $\mathcal{O}$, is expressed as: 
$$\mathcal{O}(\mathbf{X})(t)={\rm O}(\mathbf{X})(t),$$
where ${\rm O}(\mathbf{X})$ denotes the curve of outlyingness. Then, we can investigate the abnormal interaction by detecting the anomalies from these univariate curves. 

To apply the above algorithms, the users should specify the types of outliers that are of interest to detect based on the practical background, and hence determine the corresponding sequence of transformations, which naturally gives an end to these recursive procedures. According to our numerical results, we suggest applying the series of transformations $\mathcal{T}_0$, $\mathcal{T}_1$, and $\mathcal{T}_2$ in the first stage of exploratory analysis, which could handle most of the realistic functional outliers discussed and classified by \cite{hubert2015multivariate} and \cite{ArribasRomo2015discussion}. 
Changing the order of this sequence may lead to different detection results.
For example, applying $\mathcal{T}_2$ before $\mathcal{T}_1$ could turn a magnitude outlier into a shape outlier or even a non-outlying sample. 
The current order provides more interpretable results than the alternatives.
For some other combinations, e.g., $\mathcal{T}_0$, $\mathcal{T}_1$, and $\mathcal{D}_1$, it causes nearly no differences to the detection results if we reverse their order.

\vskip 12pt
\section{Simulation Study Design}
We conduct simulation studies with the following two purposes in mind: to find a proper depth notion for the functional boxplot and to assess the possible improvement in outlier detection gained by the curve transformation.
First, we introduce the investigated depth notions and the representative models contaminated with typical shape outliers. Then, we present the two scenarios used in the numerical experiments: contamination by one single shape outlier and contamination by multiple shape outliers. 

\vskip 5pt
\subsection{Existing Depth Notions}
There exist various depth notions for ranking functional data in the literature; see \citet{nieto2016topologically} and the references therein for more details. 
We investigate the following representative functional depth notions, which are sensitive to shape outliers, to search for the proper notion to describe the centrality of curves when constructing the functional boxplot.

\vskip 2pt
\subsubsection{Modified Band Depth}
\citet{lopez2009concept} proposed the idea of band depth, where the curves are ranked according to the number of envelopes formed by a fixed number of curves in the dataset, which completely contains each curve. The band depth may lead to multiple ties or even a degenerate distribution of depth values \citep{chakraborty2014data}. So, \citet{lopez2009concept} provided the modified band depth as an alternative, which is a special case of the Fraiman-Muniz depth \citep{fraiman2001trimmed}. The version based on two-curve bands is most commonly used in the literature, and it can be rapidly calculated with the following simple form \citep{sun2012exact}:
\begin{eqnarray*}
{\rm MBD}^{(2)}_n(X)=\lambda(\mathcal{I})^{-1}\int_{t\in \mathcal{I}} {2\left\{nR_i(t)-R_i^2(t)+R_i(t)-1\right\}\over n(n+1)}{\rm d}t,
\end{eqnarray*}
where $R_i(t)$ is the rank of $X_i(t)$ in the set $\{X_1(t),\dots,X_n(t)\}$, and $\lambda(\mathcal{I})$ denotes the Lebesgue measure on $\mathcal{I}$. From this perspective, the MBD of $X$ is determined by its rank at each design point.
\vskip 2pt
\subsubsection{${J}$th-order Integrated Depth}
\citet{nagy2017depth} defined the ${J}$th-order integrated depth as 
\begin{eqnarray*}
{\rm FD}_{J}(X,F_X)=\int_{0}^{1}\cdots\int_{0}^1 D\left\{(X(t_1),\dots,X(t_J)),F_{(X(t_1),\dots,X(t_J))^{\rm T}}\right\}{\rm d}t_J\cdots{\rm d}t_1,
\end{eqnarray*}
where $F_{(X(t_1),\dots,X(t_J))^{\rm T}}$ denotes the joint distribution of $(X(t_1),\dots,X(t_J))$, $(t_1,\dots,t_J)\in [0,1]^{J}$ such that $0\le t_J\le \cdots\le t_1\le 1$, and $D$ is the multivariate data depth notion decided by the users. Besides, the $J$th-order infimal depth was defined to take the minimum value of $D\left\{(X(t_1),\dots,X(t_J)),F_{(X(t_1),\dots,X(t_J))^{\rm T}}\right\}$ instead of the average. Since the infimal depth suffers from the generation of multiple ties in the ranking result, we only consider the integrated depth in this paper. Specifically, we consider the integrated depth with an order of 2, denoted by ${\rm FD}_2$, in our numerical studies.

\vskip 2pt
\subsubsection{ $L^{\infty}$ Depth}
\citet{long2015study} defined the $L^{\infty}$ depth for functional data by generalizing the $L^{p}$ depth of multivariate data \citep{zuo2000general}. Specifically, for $X\in F_X$, the $L^{\infty}$ depth has the form
\begin{eqnarray*}
{\rm L^{\infty}D}(X,F_X)=\{1+\mathbb{E}\|X-{\tilde X}\|_{\infty}\}^{-1},
\end{eqnarray*}
where $\|X\|_{\infty}=\sup_{t\in \mathcal{I}}|X(t)|$. The $L^{\infty}$ depth depends on the average distance between $X$ and ${\tilde X}\in F_X$.

\vskip 2pt
\subsubsection{Functional Directional Outlyingness}
Functional directional outlyingness \citep{dai2019directional} is a measure that accounts for the direction of an underlying observation's point-wise deviation from the bulk of a dataset, thereby revealing both the magnitude and the shape of that observation's outlyingness.
Concretely, \citet{dai2019directional} defined directional outlyingness for point-wise data as
$$\mathbf{O}(\mathbf{Y},F_{\mathbf{\textbf{Y}}})= {\rm SDO}(\mathbf{Y},F_{\mathbf{\textbf{Y}}})\cdot \mathbf{v},$$
where $F_{\mathbf{\textbf{Y}}}$ denotes the distribution of a random vector $\mathbf{\textbf{Y}}$, and $\mathbf{v}$ is the unit vector pointing from the deepest point of $F_{\mathbf{Y}}$ to $\mathbf{Y}$. The Stahel-Donoho outlyingness (SDO) \citep{stahel1981breakdown,don0ho_1982breakdown} has the form
$${\rm SDO}(\mathbf{X}(t),F_{\mathbf{X}(t)})=\sup_{\|\mathbf{u}\|=1}{\|\mathbf{u}^{\rm T}\mathbf{X}(t)-{\rm median}(\mathbf{u}^{\rm T}\mathbf{X}(t))\|\over {\rm MAD}(\mathbf{u}^{\rm T}\mathbf{X}(t))},$$
where $\mathbf{u}$ is a unit vector and MAD denotes the median absolute deviation.
Then, another two quantities are defined to measure the magnitude and shape outlyingness, respectively, of a curve
\begin{eqnarray*}
	\mathbf{MO}(\mathbf{X},F_{\mathbf{\textbf{X}}})=\int_{\mathcal{I}}\mathbf{O}(t){\rm d}t\quad{\rm and}\quad{\rm VO}(\mathbf{X},F_{\mathbf{\textbf{X}}})=\int_{\mathcal{I}}\{\mathbf{O}(t)-\mathbf{MO}\}^{\rm T}\{\mathbf{O}(t)-\mathbf{MO}\}{\rm d}t.
\end{eqnarray*}
A robust Mahalanobis distance (RMD) is calculated for each pair of $(\mathbf{MO}^{\rm T},{\rm VO})^{\rm T}$, where the covariance matrix is estimated by the minimum covariance determinant (MCD) estimator \citep{rousseeuw1985multivariate}. RMD can be treated as a two-step outlyingness and, hence, used to sort a group of functional data from the center outward.

\vskip 2pt
\subsubsection{Extreme Rank Length Depth}	
The idea of the extreme rank length depth (ERLD) or extremal depth was independently introduced by \cite{myllymaki2017} and \cite{narisetty2016extremal}. Whereas \citet{narisetty2016extremal} concentrated on the theoretical depth properties and the functional confidence intervals, \cite{myllymaki2017} focused on Monte Carlo testing based on this functional depth. 
For a group of $X_i$ discretely observed on the common design points $t_1,\dots,t_m$, $X_{ij}$ denotes the $i$th observation on the $j$-th design point.
Let $r_{1j}, r_{2j}, \dots, r_{mj}$ be the raw ranks of $X_{1j}, X_{2j}, \dots, X_{mj}$, such that the smallest $X_{ij}$ has rank 1. In the case of ties, the raw ranks are averaged. The resulting pointwise ranks are calculated as
\begin{equation}\label{eq4}
R_{ij}=\begin{cases}
r_{ij}, &\text{one-sided test, small value is considered extreme},\\
n+1-r_{ij}, &\text{one-sided test, large value is considered extreme},\\
\min(r_{ij}, n+1-r_{ij}), &\text{two-sided test}.
\end{cases}
\end{equation}
Consider the vectors of pointwise ordered ranks $\mathbf{R}_i=(R_{i[1]}, R_{i[2]}, \dots , R_{i[m]})^{\rm T}$, where \\ $\{R_{i[1]},  \dots , R_{i[m]}\}=\{R_{i1}, \dots, R_{im}\}$ and $R_{i[j]} \leq R_{i[j^\prime]}$ whenever $j \leq j^\prime$.
The ERLD of the vector $\mathbf{R}_i$ is equal to
\begin{equation*}
\label{eq:lexicalrank}
{{\rm ERLD}_i} = \frac{1}{n}\sum_{i'=1}^{n} I(\mathbf{R}_{i^\prime} \prec \mathbf{R}_i),
\end{equation*}
where $I$ denotes the indicator function and
\[
\mathbf{R}_{i^\prime} \prec \mathbf{R}_{i} \quad \Longleftrightarrow\quad
\exists\, d\leq m: R_{i[j]} = R_{i^\prime[j]}\quad \forall j < d,\  R_{i^\prime[d]} < R_{i[d]}.
\]
Simply speaking, ERLD is the left-tail stochastic ordering of the depth distributions. It

\vskip 2pt
\subsubsection{Directional Quantile}
\noindent
The precision of ERLD for finding the most extremal functions can be affected by ties which appear when $m$ and $n$ are both small. To address this drawback, \citet{myllymaki2017} gave an approximation, the directional quantile (DQ), of the two-sided ERLD as
\begin{equation}\label{Dinfty_qdir}
{\rm DQ}_i = \max_{j} \left( I(X_{ij} \geq X_{.j}) \frac{ X_{ij} - X_{.j} }{ |\overline X_{.j} - X_{.j}| } + I(X_{ij} < X_{.j}) \frac{ X_{ij} - X_{.j} }{ |\underline X_{.j}-X_{.j}| } \right),
\end{equation}
where $X_{.j}$ is the pointwise mean, and $\overline X_{.j}$ and $\underline X_{.j}$ denote the point-wise 2.5$\%$ upper and
lower quantiles, respectively, of the distribution of $X$ at the design point $t_j$. The quantities $X_{.j}$,  $\overline X_{.j}$, and $\underline X_{.j}$ are usually estimated from the observed values if they are not known analytically. 
The one-sided DQ can be defined similarly according to (\ref{eq4}) and (\ref{Dinfty_qdir}).
Essentially, ${\rm DQ}$ is the largest pointwise outlyingness of the observation $X_i(t)$.

%\vskip 5pt
%\noindent
%{\bf Binded Functional Depth}\\
%\noindent
%Except for applying each depth notion separately, we also investigated a combination of them. Concretely, we first get the rank of a curve using each depth notion, bind these ranks as a six-dimensional vector, order all the resultant vectors using the one-sided ${\rm DQ}$, and eventually obtain an overall ranking result. We denote this method as binded functional depth, denoted as ${\rm BFD}$. 
%

\subsubsection{Depth/Outlyingness Taxonomy}
${\rm FD}_2$ is the second-order extension of ${\rm MBD}$, and both are integrated notions based on pointwise ranks. As mentioned above, ${\rm RMD}$ is a two-step functional outlyingness that uses the information about distance. ${\rm ERLD}$ is based on the left-tail stochastic ordering of pointwise ranks, and ${\rm L^{\infty}D}$ and ${\rm DQ}$ rely on the maximum pointwise (scaled) distance. In practice, if both the sample size $n$ and the number of design points $m$ are small, then the rank-based depth notions suffer from a large number of ties, whereas the distance-based notion produces nearly zero ties regardless of the sizes of $n$ and $m$. ${\rm RMD}$ applies to both univariate and multivariate functional data; the other five depth notions are applicable only to univariate functional data.

\subsection{Shape Outlier Models}

\begin{figure}[!b]
	\centering
	\includegraphics[width=\textwidth]{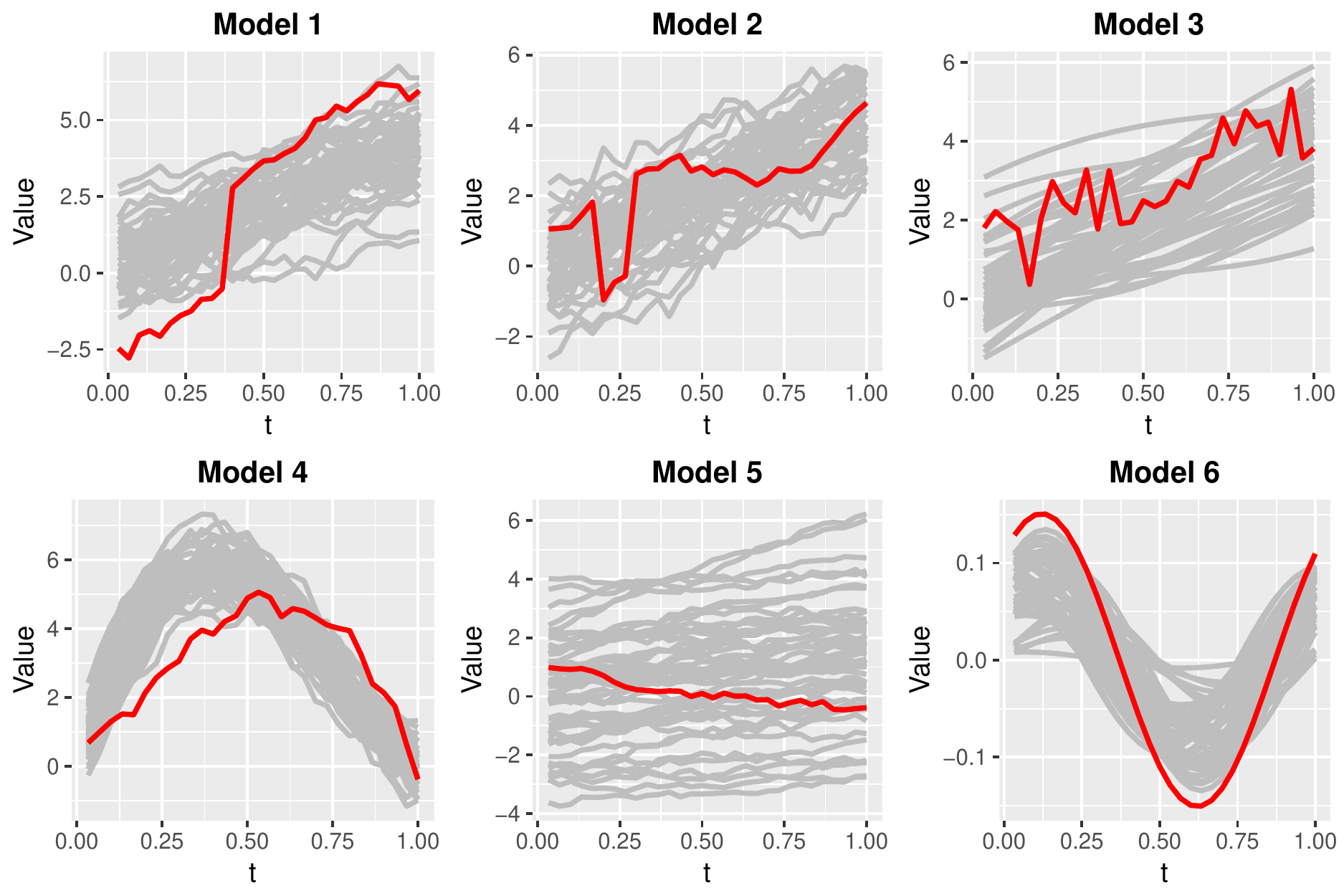}\\
	\caption{Realizations of the six investigated models contaminated by a single outlier. Grey curves: non-outlying; red curves: outlying.}
	\label{settings}
\end{figure}

We consider six types of shape outliers to comprehensively assess the performance of various notions in handling shape outliers. The models are described below:\\
\noindent
{\bf Model 0 (Clean Model)}: $X(t)=4t+e_0(t)$, $t\in[0,1]$ and $e_0(t)$ is a centered Gaussian process with the covariance function $\gamma_0(s,t)={\rm cov}\{e_0(s),e_0(t)\}=\exp{(-|s-t|)}$.\\
{\bf Model 1 (Jump)}: Main model: Model 0; contaminating model: $X(t)=4t+3I(t>U)+e_0(t)$, where $U$ follows a uniform distribution on $[0,1]$ and $I$ is an indicator function.\\
{\bf Model 2 (Peak)}: Main model: Model 0; contaminating model: $X(t)=4t+3I(U\le t\le U+0.04)+e_0(t)$.\\
{\bf Model 3 (Covariance Function)}: Main model: $X(t)=4t+e_1(t)$, where $e_1(t)$ is a centered Gaussian process with the covariance function $\gamma_1(s,t)={\rm cov}\{e_1(s),e_1(t)\}=\exp\{-(s-t)^2\}$; contaminating model: $X(t)=4t+{\tilde e}_1(t)$, where ${\tilde e}_1(t)$ is a centered Gaussian process with covariance function: $\tilde \gamma_1(s,t)={\rm cov}\{{\tilde e}_1(s),{\tilde e}_1(t)\}=\exp\{-(s-t)^{0.2}\}$.\\
{\bf Model 4 (Phase)}: Main model: $X(t)=30t(1-t)^{3/2}+e_2(t)$, where $e_2(t)$ is a centered Gaussian process with the covariance function $\gamma_2(s,t)=0.3\exp\{-|t-s|/0.3\}$; contaminating model: $X(t)=30(1-t)t^{3/2}+e_2(t)$.\\
{\bf Model 5 (Slope)}: Main model: $X(t)=A+B{\rm arctan}(t)+{e_3}(t)$, where $A$ follows a centered normal distribution with variance 4, $B$ follows an exponential distribution with mean 1, and $e_3(t)$ is a centered Gaussian process with the covariance function $\gamma_3(s,t)=0.1\exp\{-|t-s|/0.3\}$; contaminating model: $X(t)=1-2{\rm arctan}(t)+{e_3}(t)$.\\
{\bf Model 6 (Oscillation)}: Main model: $X(t)=U_{11}\cos(2\pi t)+U_{12}\sin(2 \pi t)$, where $U_{11}$ and $U_{12}$ independently follow a uniform distribution on $[0,0.1]$; contaminating model: $X(t)=U_{21}\cos(2\pi t)+U_{22}\sin(2 \pi t)$, where $U_{21}$ and $U_{22}$ independently follow a uniform distribution on $[0.1,0.12]$.

Models similar to Models 1 and 2 were considered by \citet{lopez2009concept} and \citet{long2015study}. We reduced the magnitude of the jump or peak to change the outlying curve to look more like a shape outlier. Model 3 was also introduced by \citet{lopez2009concept} and \citet{long2015study}, Model 4 was utilized by \citet{arribas2014shape},  Model 5 (with a larger residual variance) was proposed by \citet{nagy2017depth}, and Model 6 was considered by \citet{hyndman2010rainbow} and \citet{sun2011functional}. We provide a realization of each contaminated model in Figure \ref{settings}.

\vskip 12pt
\section{Simulation Study Results}
\vskip 5pt
\subsection{Contamination with One Shape Outlier}
First, we evaluate the depth notions in terms of the functional data ranking when one single shape outlier appears. 
Specifically, we generate 49 non-outlying curves from the main model and one outlier from the contaminating model for each case. The design points are 30 equidistant points on the interval $[0,1]$. 

To the raw curves, we apply the six depth notions and six joint depth notions computed according to Algorithm 2 with transformations $\mathcal{T}_0$, $\mathcal{D}_1$, and $\mathcal{T}_1$. We perform two types of transformations on the raw curves: one where we shift each curve towards its center, and one where we take the first-order differences. We calculate the rank of each sample with respect to the raw curves and two groups of transformed curves using the same depth notion; thus, we get a three-dimensional vector of ranks. Finally, we sort these vectors using the one-sided DQ. These methods are denoted by ${\rm MBD}_{\rm b}$, ${\rm FD}_{\rm 2,b}$, ${\rm RMD}_{\rm b}$, ${\rm L^{\infty}D}_{\rm b}$, ${\rm ERLD}_{\rm b}$, and ${\rm DQ}_{\rm b}$. In total, we evaluate 12 methods during each simulation and record the resulting ranks of the outlier. 
We present the average ranks given by each method over 500 replicates in Table~\ref{table1}. 

% latex table generated in R 3.3.3 by xtable 1.8-2 package
% Thu Feb 08 01:21:55 2018
\begin{table}[!t]
	\caption{The average ranks of the single outlier assigned by different methods. A smaller value indicates that the outlier is recognized as more outlying. Bold font indicates the best results. MBD: the modified band depth; ${\rm FD}_2$: the second-order integrated depth; ${\rm RMD}$: the directional functional outlyingness; ${\rm L}^{\infty}{\rm D}$: the ${\rm L}^{\infty}$ depth; {\rm ERLD}: the extreme rank length depth; ${\rm DQ}$: the directional quantile.}
	\label{table1}
	\centering
	\begin{tabular}{rcccccccccccc}
		\hline
		& Model 1 & Model 2  &Model 3  &Model 4  & Model 5  & Model 6  \\
		\hline
		${\rm MBD}$ & 7.47& 18.33& 23.19&  1.18 &36.50 & \textbf{1.00} \\
		${{\rm FD}_2}$ &6.74 &17.78 &18.15 &  1.08 & 12.04 &  1.915 \\
		${\rm RMD}$& 1.36& 2.49& 4.44& \textbf{1.00}& 3.51& 1.42 \\
		${\rm L^{\infty}D}$ & 1.66&  2.08&  4.06&  1.01& 20.08&  1.04 \\
		${\rm ERLD}$ &  3.04 & 7.27& 10.78&  1.02& 32.4 & \textbf{1.00} \\
		${\rm DQ}$ & 1.88 & 2.24&  7.70&  1.05& 31.2 & 1.03\\ \hline
	%	${\rm BFD}$ & 1.74 &  2.97&  6.23&  \textbf{1.00}& 13.54&  1.01 \\ \hline
		${\rm MBD}_{\rm b}$ &3.12& 13.71&  2.28&  1.01&  3.70 & \textbf{1.00} \\
		${\rm FD}_{\rm 2,b}$ & 2.72& 13.6 &  2.29&  1.01&  3.26&  1.77 \\
		${\rm RMD}_{\rm b}$ & 1.08 &1.25& 1.22 &\textbf{1.00}& 3.04 &1.38\\
		${\rm L^{\infty}D}_{\rm b}$ &  1.02 &1.02 &1.28& \textbf{1.00} &4.07& 1.04 \\
		${\rm ERLD}_{\rm b}$ & 1.79  &6.41 & 1.75 & \textbf{1.00} &2.79  &\textbf{1.00} \\
		${\rm DQ}_{\rm b}$ &\textbf{1.01} & \textbf{1.01} & \textbf{1.00} & \textbf{1.00} &\textbf{2.14}  &1.03 \\
		\hline
	\end{tabular}
\end{table}

When using only the raw curves, the distance-based notions (${\rm RMD}$, ${\rm L^{\infty}D}$, and ${\rm DQ}$) assign overall lower ranks to the outlier than the rank-based notions, which indicates that the distance-based depth notions are more effective in recognizing the single outlying function. 
Note that all the methods produce smaller ranks for Models 4 and 6, since the outlier in these two models achieves either the largest or smallest value across a large portion of the interval. 

All the methods are significantly improved by using the raw curves together with the transformed ones since the shape outlier becomes a magnitude outlier after the transformation. 
Overall, ${\rm DQ}_{\rm b}$ performs the best out of all the methods, almost always recognizing the outlier as the most extremal observation. 
Among the six contaminated models, the outlier in Model 5 is the hardest to detect for all the depth notions; 
the curve transformations are still helpful but not as ideal as with the other models because the signal of outlyingness is partially covered by random noise.

\vskip 5pt
\subsection{Contamination with Multiple Shape Outliers}
Next, we evaluate each method based on their outlier detection performance when a group of outliers contaminate the observations. The simulation settings are the same as in Section~3.3, except that the number of outliers is changed from 1 to 5. We perform the same two transformations of the raw curves. 
For the first six methods, we detect outliers using the functional boxplot constructed from the raw curves ranked by each of the depth notions. 
For the combination methods, we detect the outliers by applying the functional boxplots constructed with different depth notions to the raw and transformed curves separately, and collect all the detected outliers as the final result. The methods are denoted by ${\rm MBD}_{\rm c}$, ${\rm FD}_{\rm 2,c}$, ${\rm RMD}_{\rm c}$, ${\rm L^{\infty}D}_{\rm c}$, ${\rm ERLD}_{\rm c}$, and ${\rm DQ}_{\rm c}$.
We calculate the correct and false detection rates, $p_c$ and $p_f$, for each run. 
We define $p_c$ as the ratio of the number of correctly detected outliers over the number of true outliers, and
$p_f$ as the ratio of the number of falsely detected outliers over the number of non-outlying samples. 
The average performances from 500 replicates are presented in Table \ref{table2}. In addition to that, we provide the Rand index as an overall summary of these results in Table \ref{table_add2}.

\begin{table}[!b]
	\caption{The correct and false detection rates of different methods for the six models. Bold font indicates the best results.}
	\label{table2}
	\centering
\begin{tabular}{rcccccccccccc}
		\hline
		&  \multicolumn{2}{c}{Model 1} & \multicolumn{2}{c}{Model 2}  &\multicolumn{2}{c}{Model 3}  &\multicolumn{2}{c}{Model 4} & \multicolumn{2}{c}{Model 5} & \multicolumn{2}{c}{Model 6}  \\
		\hline
		&$p_c$&$p_f$&$p_c$&$p_f$&$p_c$&$p_f$&$p_c$&$p_f$&$p_c$&$p_f$&$p_c$&$p_f$\\
		\hline
		${\rm MBD}$  & 0.31 &0.00 &0.16&0.00 & 0.03&0.00 & 0.35&0.00 & 0.00 &0.05&0.00& 0.00 \\
		${{\rm FD}_2}$ & 0.30&0.00 & 0.17&0.00 & 0.01&0.00 & 0.35&0.00 & 0.00 &0.02&0.00& 0.00 \\
		${\rm RMD}$  & 0.23 &0.00 &0.13&0.00 & 0.01&0.00 & 0.32&0.00 & 0.00&0.01& 0.00& 0.00 \\
		${\rm L^{\infty}D}$ & 0.31&0.00 & 0.17&0.00 & 0.01&0.00 & 0.42&0.00 & 0.00 &0.03&0.00&0.00  \\
		${\rm ERLD}$  & 0.22&0.00 & 0.17&0.00 & 0.02&0.00 & 0.19&0.00 & 0.00&0.04& 0.00& 0.00 \\
		${\rm DQ}$ & 0.19&0.00 & 0.14&0.00 & 0.01&0.00  &0.18&0.00 & 0.00&0.03& 0.00& 0.00 \\ \hline
		%${\rm BFD}$ & 0.27 &0.00 &0.18&0.00 & 0.01&0.00 & 0.32&0.00 & 0.00&0.02& 0.00&0.00 \\ \hline
		${\rm MBD}_{\rm c}$ & 0.81&0.01 & 0.89&0.00 & \textbf{1.00} &0.02 &0.76 &0.00 &\textbf{0.80}&0.16 & 0.00&0.00  \\
		${\rm FD}_{\rm 2,c}$  &0.79&0.00 & 0.90 &0.01 &\textbf{1.00}&0.00 & 0.73&0.00 & 0.78&0.12 & 0.00&0.00  \\
		${\rm RMD}_{\rm c}$ & 0.99 &0.00 &\textbf{1.00} &0.00 &\textbf{1.00} &0.00 &0.73&0.00 & 0.72&0.12 & \textbf{0.01}&0.00  \\
		${\rm L^{\infty}D}_{\rm c}$ & \textbf{1.00} &0.00 &\textbf{1.00} &0.00 &\textbf{1.00}&0.01 & \textbf{0.77} &0.00 &\textbf{0.80}&0.15 & 0.00&0.00  \\
		${\rm ERLD}_{\rm c}$ & 0.86 &0.00 &0.98 &0.00 &\textbf{1.00}&0.00 & 0.50&0.00 & 0.15&0.06 & 0.00&0.00  \\
		${\rm DQ}_{\rm c}$ & \textbf{1.00}&0.00 & \textbf{1.00}&0.00 & \textbf{1.00}&0.00 & 0.51 &0.00 &0.07&0.04 & 0.00&0.00  \\
		\hline
	\end{tabular}
\end{table}

\begin{table}[!b]
	\caption{The Rand index of different methods for the six models. Bold font indicates the best results.}
	\label{table_add2}
	\centering
		\begin{tabular}{rcccccccccccc}
		\hline
		&  Model 1 & Model 2  &Model 3  &Model 4 & Model 5 & Model 6  \\
		\hline
		&RI&RI&RI&RI&RI&RI\\
		\hline
		${\rm MBD}$  & 0.87 &0.83 &0.82&0.87 & 0.74&0.82\\
		${{\rm FD}_2}$ & 0.86&0.84 & 0.82&0.87 & 0.78&0.82  \\
		${\rm RMD}$  & 0.85 &0.83 &0.82&0.87 & \textbf{0.80}&0.82  \\
		${\rm L^{\infty}D}$ & 0.87&0.84 & 0.82&0.88 & 0.76&0.82   \\
		${\rm ERLD}$  & 0.85&0.84 & 0.82&0.84 & 0.75&0.82  \\
		${\rm DQ}$ & 0.84&0.83 & 0.82&0.83 & 0.76&0.82   \\ \hline
		%${\rm BFD}$ & 0.27 &0.00 &0.18&0.00 & 0.01&0.00 & 0.32&0.00 & 0.00&0.02& 0.00&0.00 \\ \hline
		${\rm MBD}_{\rm c}$ & 0.96&0.98 & 0.96&\textbf{0.95} & 0.74 &0.82  \\
		${\rm FD}_{\rm 2,c}$  &0.96&0.98 & \textbf{1.00} &0.94 &0.76&0.82   \\
		${\rm RMD}_{\rm c}$ & 0.99 &\textbf{1.00} &\textbf{1.00} &0.94 & 0.74 &0.82  \\
		${\rm L^{\infty}D}_{\rm c}$ & \textbf{1.00} &\textbf{1.00} &\textbf{1.00} &\textbf{0.95} &0.75&0.82  \\
		${\rm ERLD}_{\rm c}$ & 0.97 &0.99 &0.98 &0.90 &0.75&0.82 \\
		${\rm DQ}_{\rm c}$ & \textbf{1.00}&\textbf{1.00} & \textbf{1.00}&0.90 & {0.77}&0.82   \\
		\hline
	\end{tabular}
\end{table}

When using only the raw curves, the correct outlier detection rates from all six models are quite poor because the functional boxplot is more sensitive to the magnitude outliers than to the shape outliers that these underlying models are mostly contaminated by. 
In Models 1, 2, and 4, the outliers sometimes reach the maximum or minimum values at some part of the design interval; therefore, they are detected with higher rates. 
In Models 3 and 5, the outliers are shape outliers deeply buried in the bulk of the dataset, and are rarely detected. 
The failure to detect the outliers in Model 6 is due to the coefficients generated from two adjacent uniform distributions, which can be viewed as the same distribution. Consequently, the level of outlyingness throughout the whole interval is never substantial enough to be recognized. 

As above, the performances of all the methods improve substantially when the raw curves are combined with the transformed curves. This indicates that the shape outliers are effectively changed into magnitude outliers by the transformations.  
Again, the distance-based depth notions perform better than the rank-based ones. Among the distance-based notions, ${\rm L}^{\infty}{\rm D}_{\rm c}$ provides the best results and ${\rm RMD}_{\rm c}$ is quite comparable. ${\rm MBD}_{\rm c}$ and ${\rm FD}_{\rm 2,c}$ perform the worst for Models 1 and 2, while ${\rm ERLD}_{\rm c}$ and ${\rm DQ}_{\rm c}$ perform the worst for Models 4 and~5. 

\subsection{Comparison of Transformations}
Finally, we make a comparison of the commonly used transformations mentioned in Section 2.3. Specifically, we consider five sets of transformations: $\{\mathcal{T}_0\}$, $\{\mathcal{T}_0, \mathcal{T}_1, \mathcal{D}_1\}$,  $\{\mathcal{T}_0, \mathcal{T}_1, \mathcal{T}_2\}$, $\{\mathcal{T}_0, \mathcal{D}_1, \mathcal{D}_2\}$, and $\{\mathcal{T}_0, \mathcal{T}_1, \mathcal{T}_2, \mathcal{D}_1, \mathcal{D}_2\}$. The comparison is carried out under the same setting as in Section 3.1 and we report the results for two depth notions, $DQ$ and $L^{\infty}$, which performed best in the previous studies. 
We summary the results in Table \ref{table_add}.

Through transformation, the average ranks are significantly reduced, as in Section 3.1. 
Among different combinations, $\{\mathcal{T}_0, \mathcal{D}_1, \mathcal{D}_2\}$ basically fails in Model 5 although it handles the first three models quite well.  
$\{\mathcal{T}_0, \mathcal{T}_1, \mathcal{D}_1\}$ provide the best performance, and $\{\mathcal{T}_0, \mathcal{T}_1, \mathcal{T}_2\}$ is quite comparable except for Model 1. Moreover, $\{\mathcal{T}_0, \mathcal{T}_1, \mathcal{T}_2\}$ lead to more interpretable classification of outliers as we discussed in Section 2.3.
$\{\mathcal{T}_0, \mathcal{T}_1, \mathcal{T}_2, \mathcal{D}_1, \mathcal{D}_2\}$ 
does not lead to significant improvement over $\{\mathcal{T}_0, \mathcal{T}_1, \mathcal{D}_1\}$ or $\{\mathcal{T}_0, \mathcal{T}_1, \mathcal{T}_2\}$. Considering its simplicity and interpretability, we recommend the sequence, $\{\mathcal{T}_0, \mathcal{T}_1, \mathcal{T}_2\}$, for the first step of functional data exploratory analysis.

\begin{table}[!t]
	\caption{The average ranks of the single outlier assigned by ${\rm DQ}$ and ${\rm L^{\infty}D}$ under different combinations of transformations. A smaller value indicates that the outlier is recognized as more outlying. Bold font indicates the best results.}
	\label{table_add}
	\centering
	\begin{tabular}{llcccccccccccc}
		\hline
	Transformations&Depth & Model 1 & Model 2  &Model 3  &Model 4  & Model 5  & Model 6  \\
		\hline
	\multirow{2}{*}{$\{\mathcal{T}_0\}$}	&${\rm L^{\infty}D}$ & 1.83&  2.44&  4.04&  1.01& 21.90&  \textbf{1.01} \\
		&${\rm DQ}$ & 2.17 & 3.04&  7.49&  1.04& 31.93 & 1.04\\ \hline
	\multirow{2}{*}{$\{\mathcal{T}_0, \mathcal{T}_1, \mathcal{D}_1\}$}	&${\rm L^{\infty}D}$ &  1.02 &1.01 &1.26& \textbf{1.00} &5.60& \textbf{1.01} \\
		&${\rm DQ}$ &1.01 &1.01 & \textbf{1.00} & \textbf{1.00} &\textbf{3.56}  &1.04 \\
		\hline
		\multirow{2}{*}{$\{\mathcal{T}_0, \mathcal{T}_1, \mathcal{T}_2\}$}&${\rm L^{\infty}D}$ & 1.29&  1.15&  1.41&  \textbf{1.00}& 4.04&  1.03 \\
		&${\rm DQ}$ & 1.24 & 1.14&  1.04& \textbf{1.00}& 3.69 & 1.07\\ \hline
		\multirow{2}{*}{$\{\mathcal{T}_0, \mathcal{D}_1, \mathcal{D}_2\}$}&${\rm L^{\infty}D}$ &  \textbf{1.00} &\textbf{1.00} &\textbf{1.00}& 1.15 &22.04& \textbf{1.01} \\
		&${\rm DQ}$ &\textbf{1.00} &\textbf{1.00} & \textbf{1.00} & 1.25 &16.57  &1.04 \\
		\hline
		\multirow{2}{*}{$\{\mathcal{T}_0, \mathcal{T}_1, \mathcal{T}_2, \mathcal{D}_1, \mathcal{D}_2\}$}	&${\rm L^{\infty}D}$ & 1.02&  \textbf{1.00}&  1.06&  \textbf{1.00}& 5.65&  1.02 \\
	&${\rm DQ}$ & 1.01 & \textbf{1.00}&  \textbf{1.00}&  \textbf{1.00}& 4.58 & 1.07\\ \hline
	\end{tabular}
\end{table}

Nevertheless, in the simulation studies, we did not cover all the possible cases of contamination, e.g., multivariate functional outliers or outliers due to warping. Under these cases, other transformations such as $\mathcal{O}$ and $\mathcal{R}$ should also help detect potential outliers. We address these two specific scenarios with two applications in the next section.

\vskip 12pt	
\section{Examples of Functional Outlier Detection and Taxonomy}
In this section, we assess the practical performance of changing shape outliers to magnitude outliers through data transformation in several applications. 
We find that this method not only provides a simple way to handle shape outliers, but also leads to new findings. 
\vskip 5pt
\subsection{World Population Data}
\begin{figure}[!b]
	\centering
	\includegraphics[width=10cm]{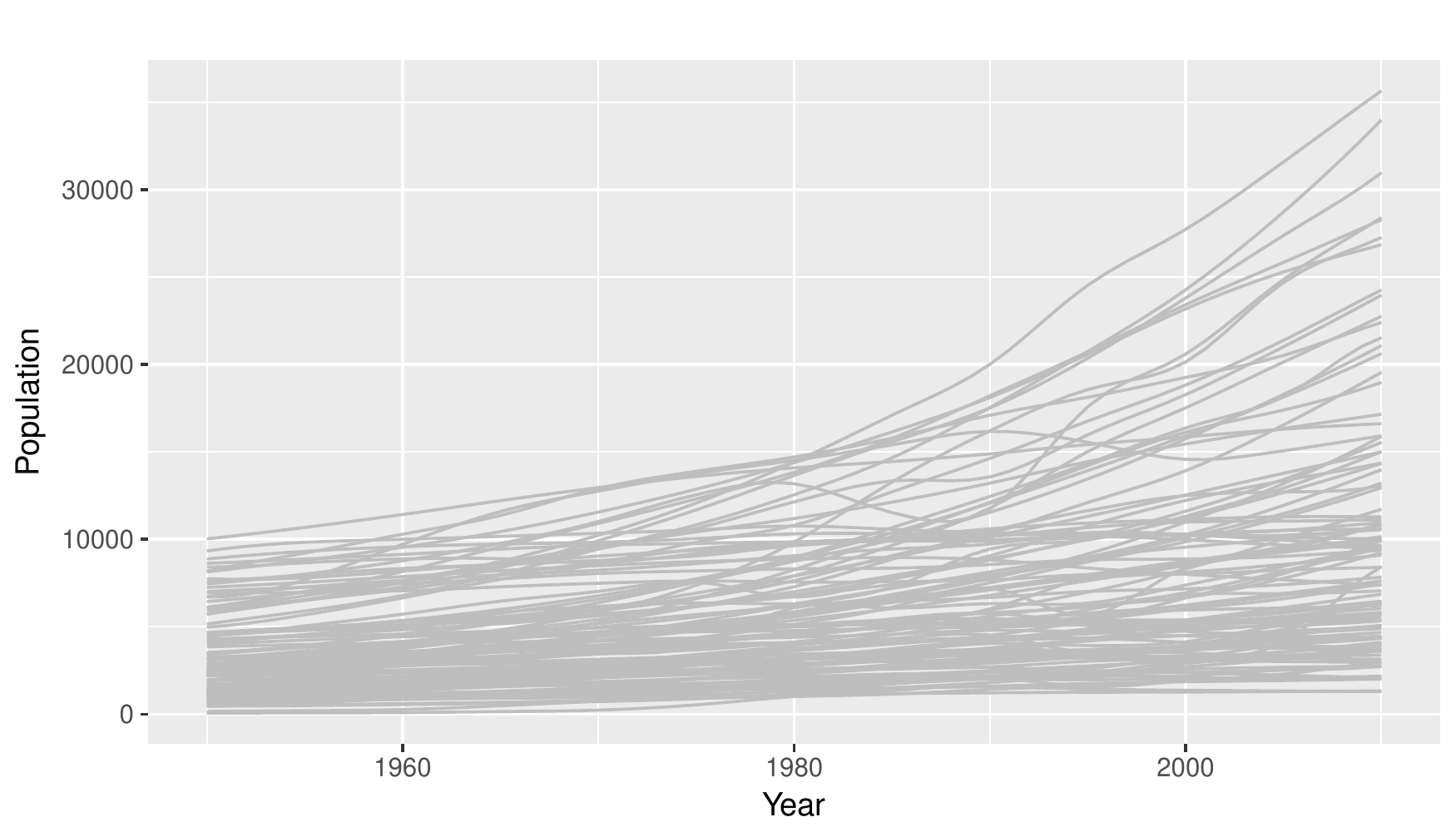}\\
	\caption{The population curves of 105 countries from 1950 to 2010.}
	\label{population_curve}
\end{figure}

First, we consider the world population data (United Nations 2016), which was analyzed by \citet{nagy2017depth}. 
This dataset (Total Population-Both Sexes) includes estimates of the total population in 233 countries, areas, or regions in July, 1950--2010. 
We follow \citet{nagy2017depth}'s preprocessing of the dataset by selecting those samples that represent populations numbering between one million and fifteen million on July 1, 1980. In total, 105 observations are included in our analysis; the curves are shown in Figure~\ref{population_curve}. 

\begin{figure}[t!]
	\centering
	\includegraphics[width=\textwidth,height=16.5cm]{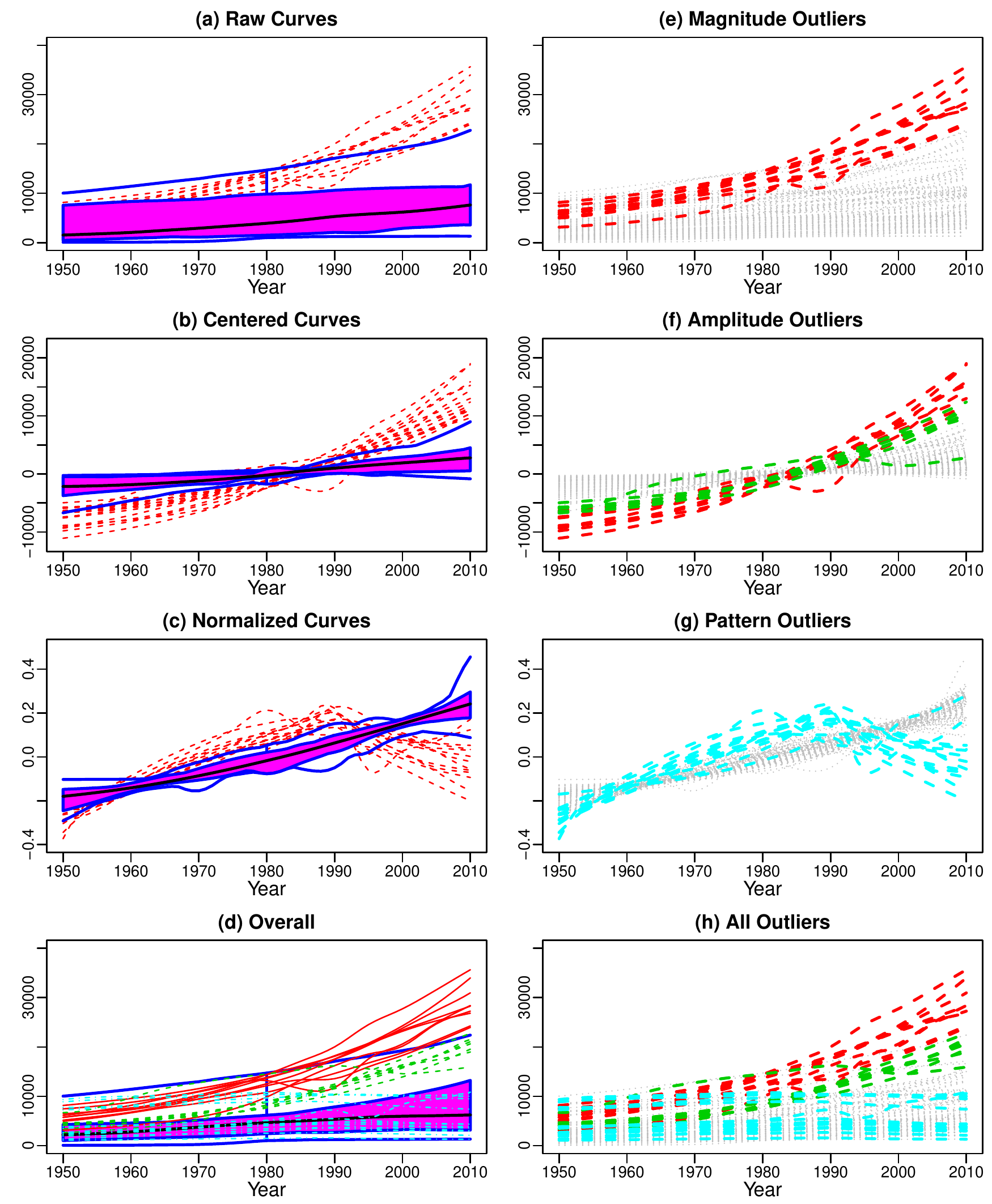}\\
	\caption{Outlier detection results from the population dataset obtained by combining the curve transformation and the functional boxplot. Left panels: the functional boxplots for the raw and transformed curves. Right panels: comparisons of outlying curves (color) with non-outlying curves (grey). First row: magnitude outliers (red) detected in the raw curves; second row: amplitude outliers (green) detected in the centered curves; third row: pattern outliers (cyan) detected in the centered and standardized curves; last row: overall description of the result.}
	\label{population}
\end{figure}

We apply Algorithm 1 with the transformations, $\mathcal{T}_0$, $\mathcal{T}_1$, and $\mathcal{T}_2$, to this dataset and construct the boxplots with the $L^{\infty}$ depth as suggested by our simulation study. The results are visualized in Figure \ref{population}, and the countries detected as outliers are provided in Table \ref{t3}.
Since the raw curves are transformed twice sequentially, the detected outliers are divided into three categories: magnitude outliers, amplitude outliers, and pattern outliers, according to our taxonomy described in Section 2.

	The magnitude outliers (see Figure \ref{population}(e)) achieve relatively large populations at the end of the investigated period. For example, the largest population, about 36 million, among the 105 countries included in our analysis is in Sudan, 2010. It was previously suggested that magnitude outliers could be detected simply with a boxplot of the means/medians of curves \citep{xie2017geometric}. However, the functional boxplot makes use of the whole curve, which is more comprehensive and, hence, captures more details about the dataset. 
	
	Amplitude outliers are curves with unusual oscillation levels. In this study, as shown in Figure \ref{population}(b) and (f), most of these countries' populations have a higher increment. Some magnitude outliers are also flagged as abnormal in terms of the amplitude, but we prefer to classify  these curves solely as magnitude outliers to get distinct sets for different categories. 
	In Figure 4(h), the green curves represent the amplitude outliers, and they are not outlying at all in terms of the magnitude. Here, the Syrian Arab Republic, which has the ninth largest population increment at about 18 million from 1950 to 2010, represents an amplitude outlier. Six other countries with higher increments than the Syrian Arab Republic have already been flagged as magnitude outliers in the first step. These countries in the above two categories are located at either Middle East or Africa, hence, share similar local economic and political environments.
	
	\begin{table}[!t]
		\caption{Outlier detection results for the population dataset using the ${\rm L}^{\infty}$ depth.}
		\label{t3}
		\centering	
		{\small
			\begin{tabular}{|C{5cm}|C{5cm}|C{5cm}|}
				\hline
				Magnitude Outliers& Amplitude Outliers & Shape Outliers\\
				\hline 
				Mozambique, Uganda, Sudan, Ghana, Afghanistan, Nepal, Malaysia, Iraq, Saudi Arabia
				&
				Madagascar, Angola, Cameroon, C{\^ o}te d'Ivoire, Kazakhstan, Syrian Arab Republic, Yemen

				&  Rwanda, Armenia, Georgia, Belarus, Bulgaria, Czech Republic, Hungary, Republic of Moldova, Estonia, Latvia, Lithuania, Bosnia and Herzegovina, Croatia \\ 
				\hline 
			\end{tabular}} 
		\end{table}
	
	The curves detected as outliers in the final step are called pattern outliers because they reveal significantly different patterns relative to the rest of the dataset after centering and normalizing. As shown in Figure \ref{population}(g), most pattern outliers (cyan) achieve peaks during 1980--1990,  and drop rapidly afterwards. All these countries, except for Rwanda, are located in Eastern Europe and share some common historical and economic background. 
	
	From Figure \ref{population}(h), it is quite difficult to locate the anomalies in the amplitude outliers (green) or the pattern outliers (cyan). However, these anomalous curves are turned into magnitude outliers by the transformations. Also, our taxonomy interprets the detected outliers well. Our procedure extracts much more information from the dataset than that of \citet{nagy2017depth}.

	\vskip 5pt
	\subsection{Annual Sea Surface Temperature Data}
	Sea surface temperature (SST) data can be utilized to monitor El Ni{\~ n}o phenomena, a fundamental measure of global climate change. Such data have been analyzed by \citet{hyndman2010rainbow}, \citet{sun2011functional}, and \citet{xie2017geometric}. We consider the dataset used by \citet{xie2017geometric} from the\href{http://www.cpc.ncep.noaa.gov/data/indices/ersst3b.nino.mth.81-10.ascii}{ Climate Prediction Center}. The dataset consists of observations from multiple regions, January 1950 to December 2014; we focus on the records from the Ni{\~ n}o 1+2 regions.

		\begin{figure}[!t]
			\centering
			\includegraphics[width=\textwidth]{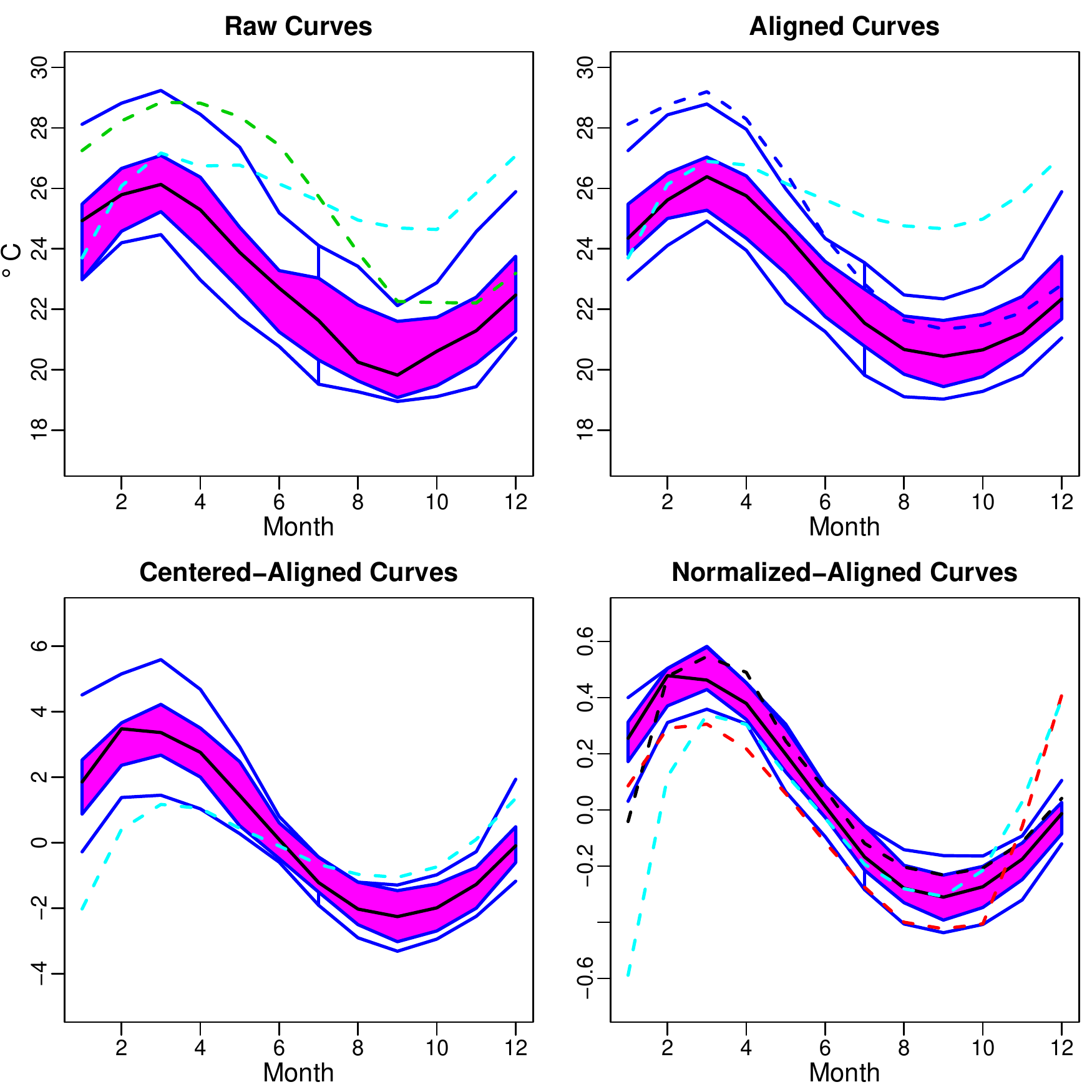}\\
			\caption{The functional boxplots constructed using the raw and three transformed SST curves. From left to right: functional boxplots based on the raw curves, the aligned curves, the centered-aligned curves, and the normalized-centralized-aligned curves. Detected outliers are presented as dashed curves: 1957 (black), 1982 (red), 1983 (green), 1997 (cyan), and 1998 (blue).}
			\label{SST_fbplot}
		\end{figure}

			\begin{figure}[!t]
				\centering
				\includegraphics[width=\textwidth,height=17cm]{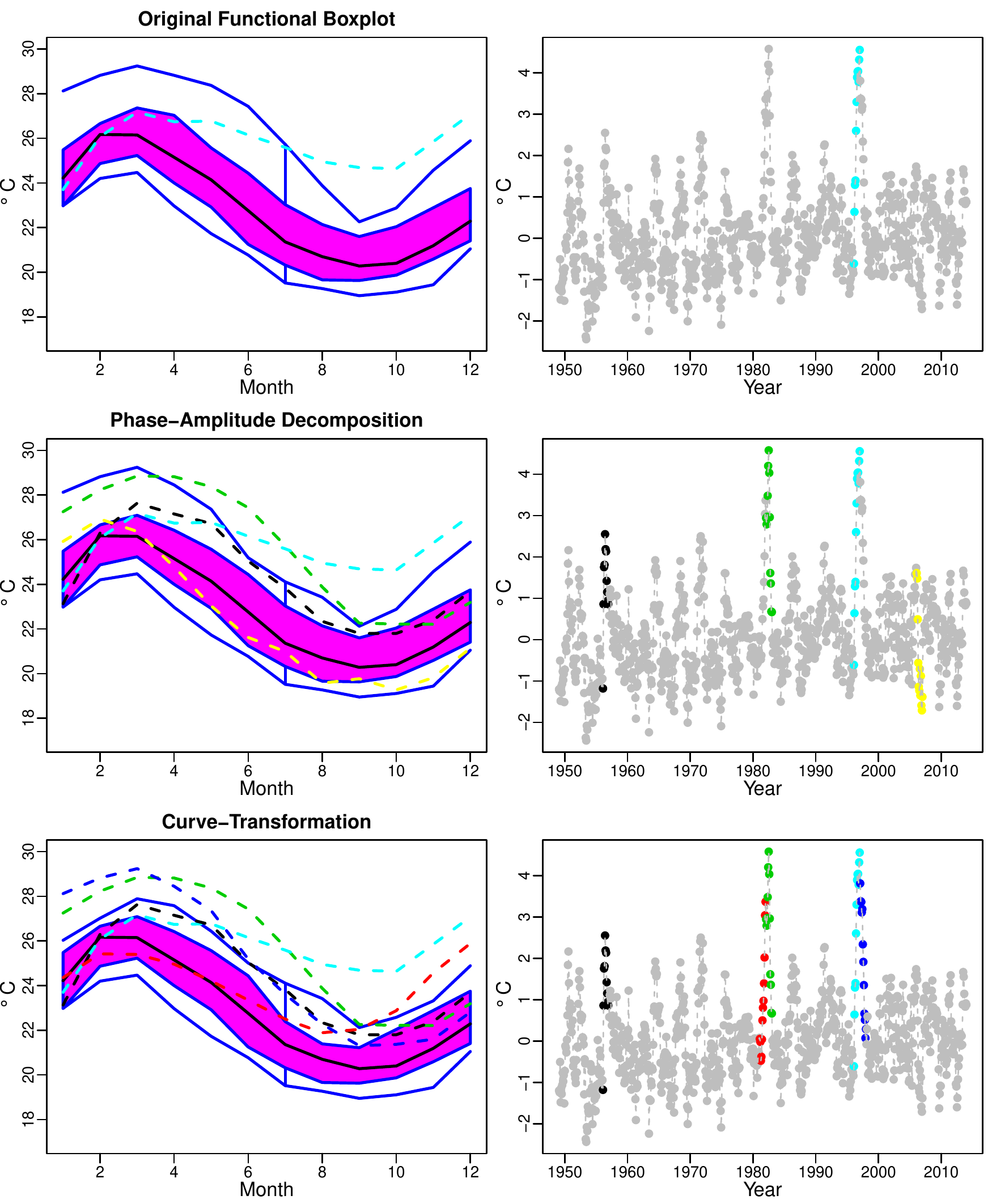}\\
				\caption{Left panels: outlier detection results from the sea surface temperature dataset using different methods: the original functional boxplot (top), phase-amplitude decomposition (middle), and a combination of the functional boxplot and curve transformation (bottom). Right panels: the detected outliers' locations in the plot of annual sea surface temperature anomalies. Dashed curves: 1957 (black), 1982 (red), 1983 (green), 1997 (cyan), 1998 (blue), and 2007 (yellow).}
				\label{Alignment}
			\end{figure}
			
    We applied Algorithm 1 with transformations $\mathcal{T}_0$, $\mathcal{R}$, $\mathcal{T}_1$, and $\mathcal{T}_2$ to the dataset. 
	Note that we align the curves in the first step, since \citet{xie2017geometric} showed that this dataset contains natural phase variability.
    After three types of transformations, we obtain four groups of curves. Then, we apply the functional boxplot based on the $L^{\infty}$ depth to each group, and combine the detected outliers as the final result. 
	The functional boxplots constructed using the four groups of curves are illustrated in Figure~\ref{SST_fbplot}. 
	In the first plot, 1983 and 1997 are detected as outliers because they achieve the highest temperatures during several months. 
	Specifically, 1983 provided the highest records for April to June, and 1997 provided the highest records for July to December. After alignment, 1998 turned out to be the warmest year from January to May and, hence, is detected as an outlier in the second plot. In the last plot, 1957 was outlying due to the sharp temperature increase from January to February. 1982 was also outlying due to the rapid increase from October to December.

	Figure~\ref{Alignment} shows our result, as well as the outliers detected by the other two methods, the original functional boxplot based on the modified band depth \citep{sun2011functional} and the phase-amplitude decomposition \citep{xie2017geometric}. According to a National Climatic Data Center report, two of the strongest El Ni{\~ n}o events happened during 1982--1983 and 1997--1998, which are completely detected by our methods but not by the two alternative methods. After those fours years, 1957 achieves the next highest temperatures. This is because we used the $L^{\infty}$ depth to construct the functional boxplot, and this depth notion puts more weight on extremal events, which matches well with El Ni{\~ n}o studies.

	\subsection{Global Envelope Test for Spatial Point Processes}

    The features of spatial point processes are usually summarized by a function of distance, $r$. The most commonly used characteristic of point processes is the centralized $L$-function, which is the transformation of Ripley's $K$-function \citep{IllianEtal2008}. \cite{myllymaki2017} proposed a global envelope test to assess the goodness-of-fit of point process models. Specifically, they assumed that a group of curves follow an identical distribution, e.g., $L$-functions of simulations from the same spatial point process model, and then they constructed the global envelope with the curves sorted according to ERLD or DQ. This envelope test provides not only an exact $p$-value, but also a graphical interpretation of the reason for rejection. 
           		\begin{figure}[!t]
           			\label{F:GP}
           			\centerline{\includegraphics[width=\textwidth]{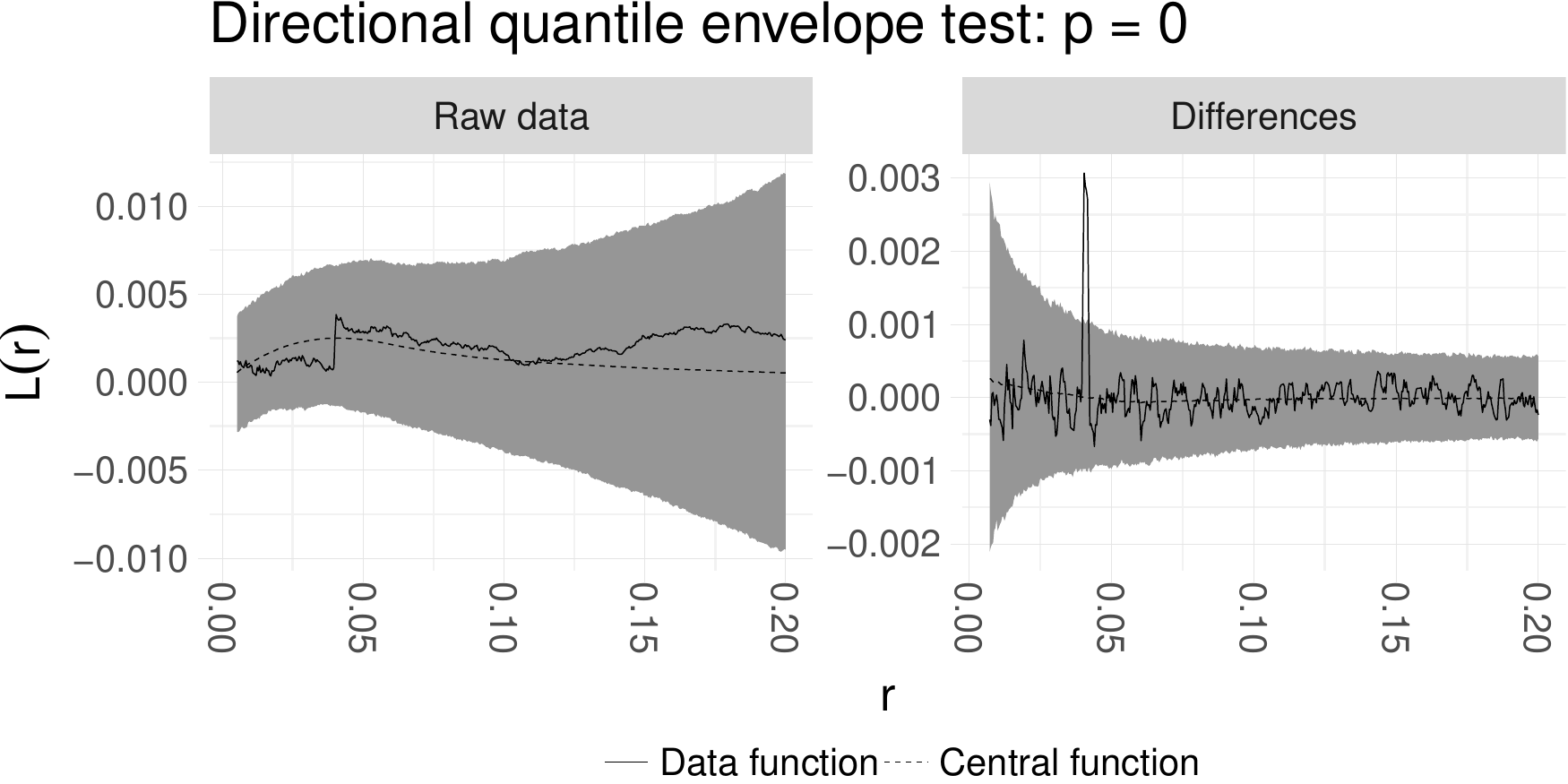}}
           			\caption{\label{F:GP} The global envelope test based on the directional quantile of the Gauss Poisson model against the Mat{\' e}rn cluster model. The extreme rank length depth $p$-value of the MC test is 0.0005. The shaded area is the $95\%$ global envelope.}
           		\end{figure}

    The differences among the $L$-functions are usually represented by the magnitude anomalies but there are functions that differ only in shape. We consider the Gaussian-Poisson model ($H_1$) with parameters $\kappa = 400, r_0=0.04, p_2 =0.2$, where $\kappa$ is the intensity of the Poisson process of the cluster centers, $r_0$ is the diameter of each cluster that consists of exactly two points, and $p_2$ is the probability that a cluster contains exactly two points. The $L$-function of this model contains a jump in the distance $r_0$. 
    We test whether this model is a Mat{\' e}rn cluster process ($H_0$). 
    In our example, we simulate the point process under $H_1$ in an area $[0, 1]^2$, and compute its $L$-function $L_0(r)$. We calculate all the functions in this section at 500 equally spaced design points. The parameters of the tested $H_0$ model are estimated and $s$ simulations of point processes are drawn from $H_0$. We set $s=1999$ . The associated $L_1(r), \ldots , L_s(r)$ are computed. Further, we choose the directional quantile (DQ) to construct the envelope as suggested by the simulation results in Section 4.1. We compute DQ for every $L$-function and apply the Monte Carlo test at a significance level of 0.05 in order to check if the chosen depth distinguishes $L_0$ as an outlier or not. One realization of the $L$-function from the $H_1$ model, together with the $95\%$ global envelopes of the null model, is shown in Figure \ref{F:GP}. We repeat these procedures 500 times and record the ratio of the positively detected outliers. 
    
    As shown in Figure~\ref{F:GP}, we apply Algorithm 2 with transformations $\mathcal{D}_0$ and $\mathcal{D}_1$. Using only the raw curves, the rejection ratio is zero. 
However, the jump anomaly in the tested curve is clearly observed after taking the first-order differences of the raw curves; the second plot in Figure \ref{F:GP}. 
Thus, the ratio was greatly improved to $100\%$ when we apply the global envelope test to the bound raw curves and the first-order differences. This confirms that the data transformation indeed improves the spatial point process test by providing more comprehensive perspectives about the data.

	\subsection{Multivariate Weather Data}
		
		We use a Spanish weather dataset from the R package {\em fda.usc} to demonstrate the curve-transfromation analysis of multivariate functional data. This dataset contains averaged daily temperature, log precipitation, and wind speed records from 1980 to 2009 at 73 weather stations in Spain. 
		The three-dimensional coordinates, longitude, latitude, and altitude, of these stations are also provided.
		The raw data are discretely observed and have been smoothed with 11 order-4 B-spline basis functions.
		
\begin{figure}[!t]
			\centering
			\includegraphics[width=\textwidth]{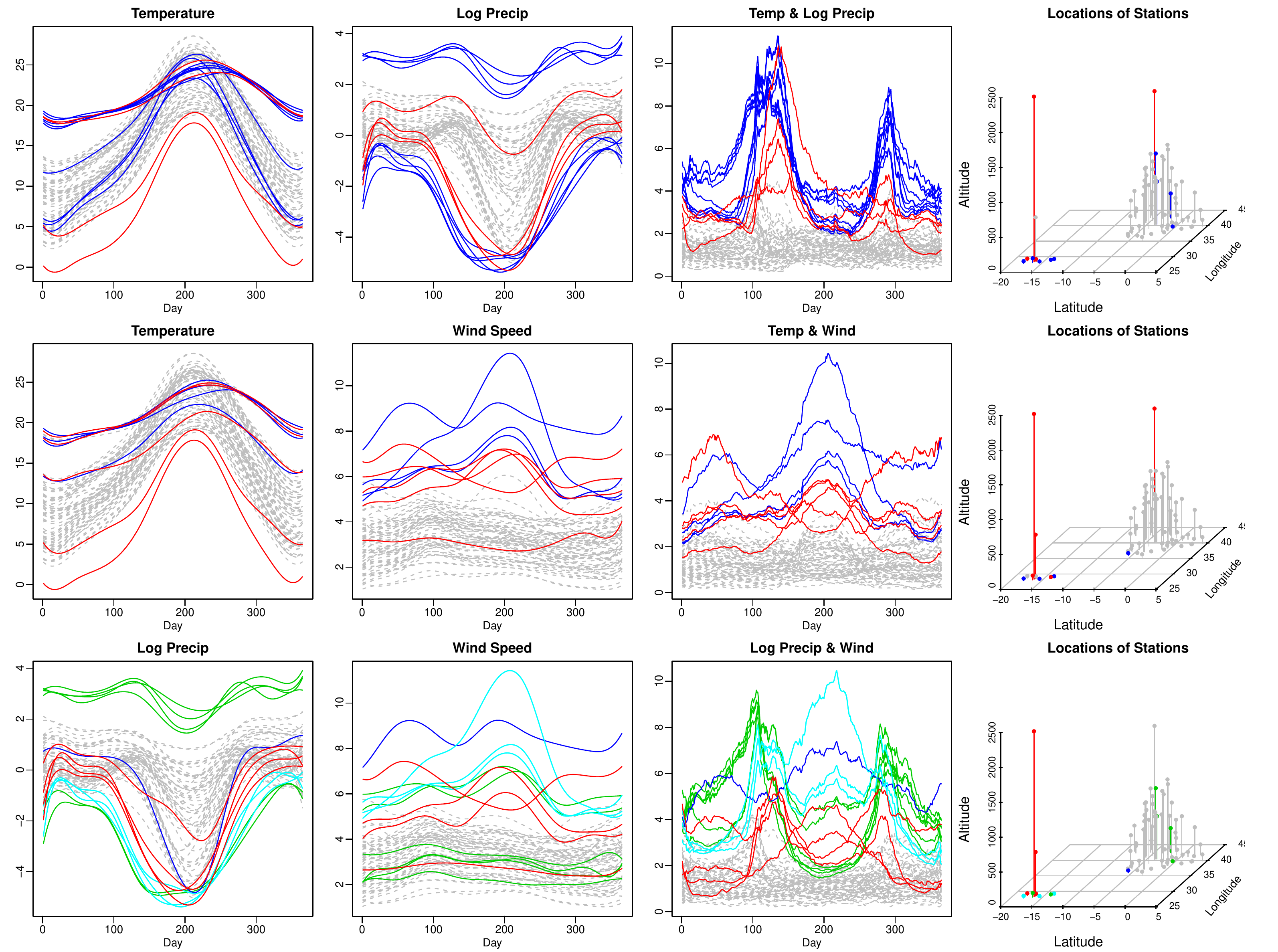}\\
			\caption{Outlier detection results from the Spain weather dataset obtained by combining the curve transformation and the functional boxplot. Each row represents a bivariate combination of the three indexes.
			First column: the first index; second column: the second index; third column: the bivariate combinations; fourth column: the locations of the stations. 
			Green curves: outliers detected in the first index and the combination; blue curves: outliers detected in the second index and the combinations; red curves: outliers detected only by the combinations; cyan curves: outliers detected by all three cases.}
			\label{Multi}
		\end{figure}

		Our goal here is to find those stations that reveal significantly different weather patterns from the majority and, further, to identify the reasons behind their anomalies. 
		We apply a functional boxplot with RMD to each type of curves to detect the marginal outliers. 
		Since we are also interested in the potential joint outliers that are outlying not for any single marginal index but for some combination of marginal indexes. 
		We apply Algorithm 1 with transformations, $\mathcal{T}_0$ and $\mathcal{O}$. 
		We calculate the pointwise SDO of the bivariate curves from each combination and get a group of univariate curves with the outlyingness as responses. 		
		Next, we detect the joint outliers using these outlyingness curves. 
		However, unlike the common case where both the remarkably small and large values are treated as anomalies, here only the larger values of the outlyingness curve are considered abnormal. Thus, we use the one-sided DQ to rank the curves from the bottom up. 
		Unlike the two-stage functional boxplot \citep{dai2018mfbplot} that detects the joint anomalies using vectors of descriptive statistics, this proposed procedure utilizes the whole curves of outlyingness and provides more concrete explanations for the detected anomalies. 
		The detection results from the three marginal and combinational cases are illustrated in Figure \ref{Multi}.

		In the first row of Figure \ref{Multi}, the magnitude outlyingness in one index helps to identify the possible shape outlyingness in the other index. 
		Specifically, the blue curves in the bottom of the log precipitation plot are detected as magnitude outliers.
		Referring to the locations of the stations, we find that these curves are recorded on the Canary Islands, which are far away from the mainland Spain. 
		At the stations with low altitudes in this area, the winter is warmer and the annual temperature variations are smaller than at most of the other stations, which means that the temperature curves are outlying in terms of shape. 
		However, these shape outliers are missed when using only the temperature curves. We manage to identify their anomalies by borrowing information from the log precipitation curves. For the other two combinations, we also obtain such benefits from the outlyingness-curves. 
		\begin{figure}[!t]
			\centering
			\includegraphics[width=\textwidth]{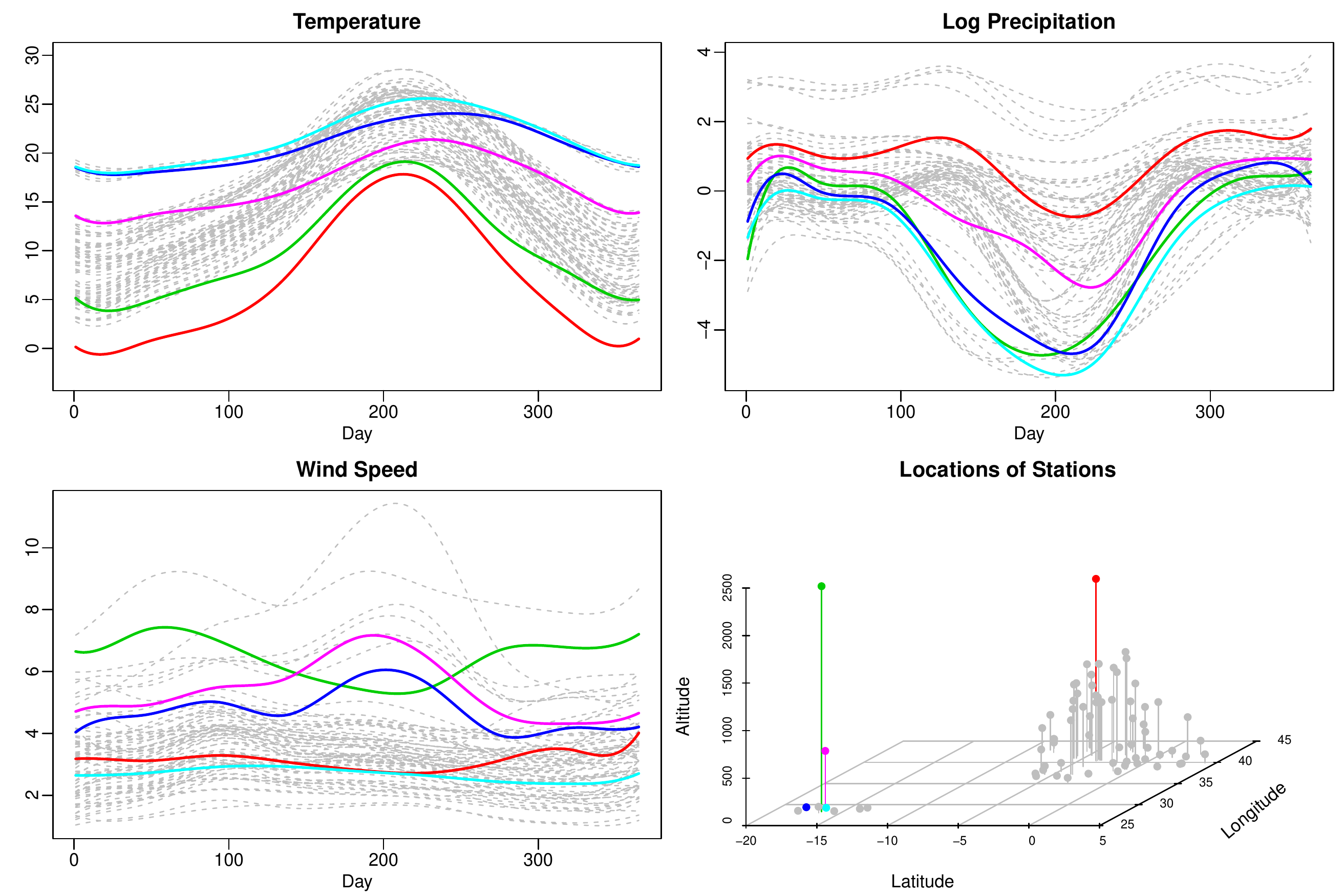}\\
			\caption{Joint outliers that are not detected as outlying by any single marginal index.} 
			\label{out_joint}
		\end{figure}	    	
    	
     In the second row of Figure \ref{Multi}, the magnitude outliers in the curves of outlyingness reveal  abnormal interactions among the marginal indexes. 
     We present in Figure \ref{out_joint} all the five joint outliers that are not identified as magnitude outliers by any marginal index. The purple one is a typical example of this category. 
     The purple weather station is located on the side of Mount Teide at an altitude of 632 meters. 
     It reveals no significant anomalies for any marginal index and, hence, we infer that its outlyingness is due to abnormal interactions among the three indexes.  
    	
	\vskip 12pt	
	\section{Conclusion}
   Turning shape outliers to magnitude outliers, which are well handled by the functional boxplot, dramatically simplifies the outlier detection procedure. Simulation studies indicate that distance-based depth notions are appropriate for constructing the functional boxplot. 
   The proposed outlier detection procedure is based on the whole curve rather than some scalars extracted from the curves. Thus, it provides more details about why a curve has been identified as an outlier.
   Applying several curve transformations sequentially provides a natural classification of the functional outliers; hence, the anomalies of these curves are easier to interpret. Data transformation also fortifies the global envelope test against more types of alternatives. As a practical suggestion, we recommend the combination of $\mathcal{T}_0$, $\mathcal{T}_1$, and $\mathcal{T}_2$ as the first step when carrying out the exploratory analysis, 
   which could handle most of the realistic functional outliers discussed and classified by \cite{hubert2015multivariate} and \cite{ArribasRomo2015discussion} as demonstrated in our numerical studies with both simulated data and real applications.

   The proposed procedure is readily extended to image or surface data, where we may replace the functional boxplot with the surface boxplot \citep{genton2014surface}.
   We have ignored possible dependencies among the trajectories for outlier detection problems throughout the current paper. For dependent functional data, the adjusted functional boxplot \citep{sun2012adjusted} with the inflating factor $F^*$ chosen by a data-driven procedure can be employed. 
Applying transformations is an intuitive and simple way to evaluate functional data from different perspectives. This is somewhat similar to measuring the dissimilarity of curves using different metrics, e.g., $L^2$, $L^{\infty}$ and semimetric of the derivatives \citep{ferraty2006nonparametric}, except that one can get the graphical interpretation of the transformed curves using our method.  
Further investigation is necessary to explore the connection between the transformations and metrics from a more theoretical point of view.

		{
			\setstretch{1.1}
			\bibliographystyle{asa}
			\bibliography{shape}
		}

	\end{document}